\documentclass[preprint,aps ,nofootinbib]{revtex4}
\usepackage{graphicx}
\pdfoutput=1
\usepackage{epsfig}
\usepackage{amsmath}
\usepackage{amsfonts}
\usepackage{amssymb}
\usepackage{color}
\usepackage{dcolumn}
\usepackage{hyperref}
\usepackage{multirow}
\usepackage{booktabs}
\usepackage{MnSymbol,wasysym}
\usepackage{braket,diagbox}
\usepackage{eurosym}
\usepackage{calrsfs}
\usepackage{subfigure}
\providecommand{\U}[1]{\protect\rule{.1in}{.1in}}

\newcommand{\f}{\begin{equation}}
\newcommand{\ff}{\end{equation}}
\newcommand{\fa}{\begin{eqnarray}}
\newcommand{\ffa}{\end{eqnarray}}

\begin{document}
\baselineskip=0.55 cm
\title{Parameter constraints on a black hole with Minkowski core through quasiperiodic oscillations}

\author{Ming-Yu Guo}
\affiliation{School of Mathematics, Physics and Statistics, Shanghai University of Engineering Science, Shanghai 201620, China}
\affiliation{Center of Application and Research of Computational Physics, Shanghai University of Engineering Science, Shanghai 201620, China}

\author{Meng-He Wu}
\email{mhwu@njtc.edu.cn} 
\affiliation{College of Physics and Electronic Information Engineering, Neijiang Normal University, Neijiang 641112, China}

\author{Xiao-Mei Kuang}
\email{xmeikuang@yzu.edu.cn}
\affiliation{ Center for Gravitation and Cosmology, College of Physical Science and Technology, Yangzhou University, Yangzhou 225002, China}

\author{Hong Guo}
\affiliation{Escola de Engenharia de Lorena, Universidade de S\~ao Paulo, 12602-810, Lorena, SP, Brazil}

\begin{abstract}
\baselineskip=0.5 cm
We investigate the geodesic motion of charged particles in the vicinity of regular black holes with a Minkowski core, embedded in a uniform magnetic field, and study the influences of magnetic field and regular black hole parameter on the radial effective potential and angular momentum of the particles' orbits. We perturb the circular orbit and analyze the characteristic frequencies of the epicyclic oscillations, which are closely related with the quasiperiodic oscillations (QPOs) phenomena of the accretion disc surrounding the black hole. Then using the MCMC simulation, we fit our theoretical results with four observational QPOs events (GRO J1655-40, XTE J1550-564, XTE J1859+226, and GRS 1915+105) and provide constraints on the magnetic field strength $B$, the characteristic radius $r$, the mass $M$, and the regular black hole parameter $g$. In particular, since the parameter $g$ describes the degree of deviation from the classical Schwarzschild black hole, our studies suggest that, within a certain level of confidence, the black holes in the current model can deviate from the classical singularity structure of Schwarzschild black holes, and exhibit quantum corrections near the core.
\end{abstract}

\maketitle
\tableofcontents

\section{Introduction}

General relativity (GR) predicts the existence of black hole, which was further verified by recent achievements in strong field regime on gravitational wave observation \cite{LIGOScientific:2016aoc,LIGOScientific:2018mvr,LIGOScientific:2020aai} and supermassive black hole shadows \cite{EventHorizonTelescope:2019dse,EventHorizonTelescope:2019ggy,EventHorizonTelescope:2019ths,
EventHorizonTelescope:2022xnr,EventHorizonTelescope:2022xqj}. However, there are many compact objects, such as wormholes and other quasars, which share some phenomena of black hole such that they are known as black hole mimickers.
Moreover, despite of many successful tests, GR itself still faces many challenges in both theoretical and observational aspects. For examples, it is still open to understand the universe expansion history, the large scale structure, singularity problem and construct a well-defined quantum gravity. Also, the uncertainties in various observational data leave some space for alternative theories of gravity generalized from GR. All these are having inspired physicists' great interest in proposing various generalized theories of gravity, which extend richer framework to further understand the nature of gravity.

It is believed that the singularity in GR can be removed by introducing a complete theory of quantum gravity. As a pioneering work \cite{Bardeen:1968} to overcome the singularity problem in a classical way, Bardeen proposed a static regular black hole. This regular black hole has an asymptotically flat center (non-singularity) instead of singularity, which
was later demonstrated to be a solution to Einstein’s theory
coupled with a magnetic monopole source \cite{Ayon-Beato:2000mjt}. After that, many regular (non-singular) black holes were constructed. Specifically, as far as we know, there are two independent ways to construct regular black holes. The first way is to directly solve equations of motion in generalized gravity theories by introducing special sources, see for examples \cite{dymnikova1992vacuum, Nicolini:2005vd,Balakin:2016mnn,Roupas:2022gee} and references therein, and the regular black holes obtained from this way behave semiclassically. The second way is to propose the regular black holes as quantum corrections to the classical black holes with singularity, see for examples \cite{Borde:1996df, Bonanno:2000ep,Gambini:2008dy,Perez:2017cmj,Brahma:2020eos}. Thus, these regular black holes in general possess quantum behaviors, in this sense, the black hole singularity could be removed or avoided by considering the quantum effects of gravity. Therefore, the regular black holes become one of the powerful tools to study the classical
limit of quantum black holes before we have a mature theory of quantum gravity. Reviews on the development of regular black holes can be seen in \cite{Torres:2022twv,Lan:2023cvz}.

An interesting topic is to study the phenomena to distinguish the regular black holes and normal black holes, especially in the strong gravity sector, the properties of which can be well reflected in the spectrum emitted by the accretion around the compact objects. Considerable radiation originates very deep in the gravitational field of these objects and usually can provide key information to test the near-horizon region of black holes. But the radiation cannot come from the central regions of black holes, so we have to rely on the radiation coming from their close surroundings, such as accretion disks \cite{Bardeen:1972fi}. The accretion disks around the central object have soft X-ray continuum emission, 
whose frequencies can help to measure the innermost disk radius, which in principle coincides with the Innermost Stable Circular Orbit (ISCO) encoding the information of the central object. Indeed, dating back to the 1970s, the authors of \cite{Syunyaev1972} proposed to study the geodesic motion in the strong field (inner disc region) using the fast variability of the X-ray flux emitted by matters close to the accreting source.
 This proposal was awakened when in the X-ray flux from accreting compact objects, people discovered the astrophysical phenomena known as Quasi Periodic Oscillations (QPOs) which have frequencies up to 450\emph{Hz}, close to those expected from bound orbits near the ISCO \cite{Lewinbook, Motta:2016vwf}. The QPOs are found by using Fourier analyses of the noisy continuous X-ray data from
 the accretion disk in (micro)quasars including black holes or neutron stars and companion stars as binary systems), and they are classified as high frequency (HF, $0.1 -1 kHz$) and low frequency (LF, $<0.1 kHz$) \cite{Stella:1997tc, Stella:1999sj}.

 HF QPOs are often observed in pairs (upper and lower frequency) and the ratios of these frequencies observed in black hole microquasars are usually around $3:2$ \cite{Kluzniak:2001ar}. It is the upper frequencies that are very close to the orbital frequencies of test particles moving on the stable circular orbit 
 located at the inner edge of the accretion disc around black holes. These oscillations can provide information about the central objects and even their origin. Several models have been proposed to describe these phenomena. Nevertheless, the first suggestion that QPOs
 may be used to test the strong-field regime of gravity was based on a simple model in which the QPO frequencies correspond to the geodesic motion of a test particle \cite{Kluzniak1990}. Due to the connection to the motion of test particles near the ISCO, properly modeling the QPO signal can provide a powerful diagnostic of strong gravitational fields.
The epicyclic motion of test particles with orbital, radial and latitudinal frequencies can be
a useful tool in modeling and explaining the observed HF QPOs, which conversely help to constrain the theoretical model parameters \cite{Bambi:2012pa,Bambi:2013fea,Maselli:2014fca,Jusufi:2020odz,Ghasemi-Nodehi:2020oiz,Chen:2021jgj,Allahyari:2021bsq,Deligianni:2021ecz,Deligianni:2021hwt,Jiang:2021ajk,Banerjee:2022chn,Liu:2023vfh,Riaz:2023yde,Rayimbaev:2023bjs,Abdulkhamidov:2024lvp,Jumaniyozov:2024eah} and references therein. On the other hand, higher precision and accuracy observations from Insight-HXMT (Hard X-ray Modulation Telescope) \cite{Lu:2019rru} and next generation X-ray time-domain Telescope Einstein Probe \cite{yuan2018einstein} 
are also expected to provide very stringent constraints on parameters of the central objects.

Here, we are interested in investigating the epicyclic motion of a charged particle and its application to the observed data of QPOs in the vicinity of a regular black hole with Minkowski core \cite{Ling:2021olm} immersed in a uniform magnetic field. It is noticed that the electromagnetic field is 
still interesting in the study of astrophysical black holes. 
Though a black hole is known to have no intrinsic magnetic field because, in GR, the gravitational collapse of massive objects will quickly decay \cite{anderson1970gravitational}, external factors, such as 
the surrounding accretion discs, 
may contribute to the presence of a magnetic field. So, it is commonly believed that black holes can be surrounded by magnetic field in an astrophysical sector, but such a magnetic field does not modify the background and can be treated to be a test field, if its strength is small enough.
However, even a small magnetic field will significantly affect the motions of charged particles, making the magnetic field increasingly important for testing the properties in the vicinity of black holes.
Thus, we will first introduce the regular black hole with Minkowski core and the uniform magnetic field, then deal with the circular orbits of the charged particle in this background. We shall evaluate the epicyclic frequencies of the oscillations around these circular orbits.

One aim of this paper is to borrow the dynamic of charged particle to further 
investigate the strong field regime of the regular black hole with Minkowski core, differentiating from Schwarzschild black hole in GR. Another aim is to provide parameters constraints on the backgrounds via the observed QPO frequencies. To this end, with the use of Markov Chain Monte Carlo (MCMC) algorithm \cite{Foreman-Mackey:2012any}, we shall fit the 
theoretical predictions for the QPO frequencies to the observational X-ray data of GRO J1655-40, XTE J1550-564, XTE J1859+226 and GRS 1915+105 \cite{Strohmayer:2001yn}, respectively, which will provide constraints on the parameters of the regular black hole with Minkowski core as well as the strength of the magnetic field.

The remaining of this paper is organized as follows. In section \ref{sec:background}, we analyze the equations of motion for a charged particle around a regular black hole with Minkowski core immersed in an external magnetic field. In section \ref{sec:EpicyclicMotion}, we 
consider the epicyclic oscillation of the particle's circular motion in the aforementioned background, and investigate 
the effects of the magnetic field and black hole parameters on the three characterized frequencies in the oscillation. In section \ref{sec:constraint} by employing MCMC method, we use the observational data in four QPOs events 
to constrain the magnetic field and model parameters. Section \ref{sec:conclusion} contributes to our conclusion and discussion.

\section{Charged test particles around the regular black hole in a uniform magnetic field}\label{sec:background}

First, we provide a brief introduction to the regular black holes with a Minkowski core, as proposed in \cite{Ling:2021olm}. The metric is given by
\begin{equation}\label{eq:1}
d s^2=-f(r) d t^2+f(r)^{-1} d r^2+r^2\left(d \theta^2+\sin ^2(\theta)\right) d \phi^2,
\end{equation}
where $f(r)=1-2m(r)/r$, with $m(r)$ having the form
\begin{equation}\label{eq:2}
m(r)=M e^{-g^n M^\gamma / r^n}.
\end{equation}
The parameter $g$ shifts the maximum Kretschmann scalar curvature to a larger radius, affecting spacetime geometry and gravitational properties. Additionally, $g$ initially increases the Hawking temperature before decreasing it, implying changes in the black hole's thermodynamics, such as entropy and heat capacity, and potentially influencing its stability and evaporation dynamics \cite{Ling:2021olm}. As noted in \cite{Ling:2021olm}, these black holes possess asymptotically non-singular Minkowski cores at their center. To ensure that Eq. \ref{eq:2} characterizes regular black holes with sub-Planckian curvature, the following constraints are necessary: $3/n<\gamma<n$ and $n\geqslant2$.
Furthermore, these regular black holes have a one-to-one correspondence with those having asymptotically de Sitter core, where the mass function $m(r)$ is defined as follows:
\begin{equation}\label{eq:3}
m(r)=\frac{M r^{\frac{n}{\gamma}}}{\left(r^n+\gamma g^n M^\gamma\right)^{1 / \gamma}},
\end{equation}
when $\gamma=1, n=3$, it is the Hayward black hole.
In this paper, we shall focus on the black hole described by Eq. \ref{eq:2} with $\gamma=1, n=3$.
This selection ensures that the black hole described by equation \ref{eq:2} exhibits a maximum Kretschmann scalar curvature that is mass-independent, as noted in \cite{Ling:2021olm}.

\begin{figure}[ht]
    \centering
    \includegraphics[width=0.4\linewidth]{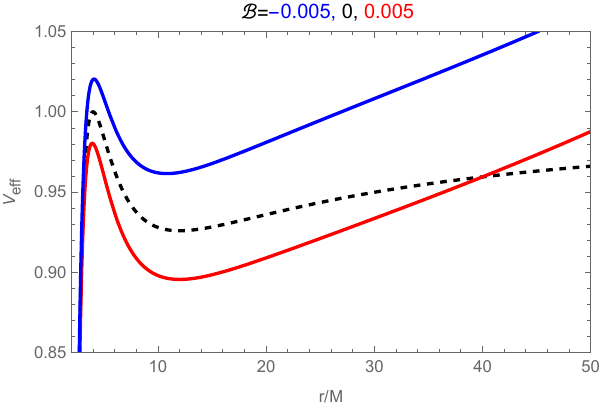}
    \hspace{1cm}
     \includegraphics[width=0.4\linewidth]{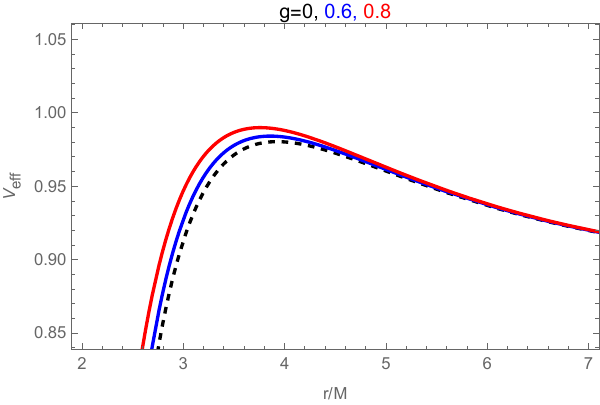}
    \caption{Radial dependence of the effective potential for the radial motion of test particles around the magnetized regular black hole where $\mathcal{L}=4$. Left panel: $V_{\text{eff}}$ plots of various combinations of $\mathcal{B}$ with parameter $g=0.1$. Right panels: $V_{\text{eff}}$ diagram of various combinations of parameter $g$ when fixed magnetic field parameter $\mathcal{B} = 0.005$.}
    \label{figa3}
\end{figure}

Next, we consider the regular black hole immersed in an external asymptotically uniform magnetic field aligned with the axis of symmetry of the black hole, with the strength of the magnetic field denoted as $B$.
Since the regular black hole we study is neutral, the electromagnetic four-vector potential $A^\alpha$ in the Lorentz gauge is represented as \cite{Abdujabbarov_2014,wald1974black}:
\begin{equation}\label{eq:4}
A^\alpha=C_\phi \xi_{(\phi)}^\alpha,
\end{equation}
where $C_\phi$ depends on the magnetic field strength $B$, and $\xi_{(\phi)}^\alpha$ is space-like axial Killing vector. The coefficients $C_\phi$ can be obtained from the Wald’s solution \cite{wald1974black} :
\begin{equation}\label{eq:5}
C_\phi=\frac{B}{2}.
\end{equation}
At the end of this section, we turn our attention to the dynamics of charged particles orbiting a regular black hole situated within an external, asymptotically uniform magnetic field. The Hamiltonian for the charged particle can be expressed as
\begin{equation}
H=\frac{1}{2} g^{\mu \nu}\left(\frac{\partial S}{\partial x^\mu}-q A_\mu\right)\left(\frac{\partial S}{\partial x^\nu}-q A_v\right),\label{eq:6}
\end{equation}
where $S$ is the Jacobi action, $q$ is the electric charge of the particle. 
Given that the metric in Eq. \ref{eq:1} and the electromagnetic four-vector potential in Eq. \ref{eq:4} are independent of the coordinates $(t, \phi)$, this leads to the conservation of two quantities: the energy $E$ and angular momentum $L$. The system under consideration is integrable, allowing the Jacobi action $S$ to be expressed as 
\begin{equation}
S = \frac{1}{2} m^2 \tau  -Et + L\phi + S_r + S_\theta,\label{eq:7}
\end{equation}
where $S_r$ and $S_\theta$ are 
the radial and angular functions of $r$ and $\theta$, and $\tau$ and $m$ are the proper time and mass of the particle, respectively. The Hamilton-Jacobi equation for a charged particle is given by
\begin{equation}
\mathcal{H}=\frac{\partial S}{\partial \tau}+H=0.\label{eq:8}
\end{equation}
By substituting Eq. \ref{eq:6} and \ref{eq:7} into Eq. \ref{eq:8}, the Hamilton-Jacobi equation can be rewritten as
\begin{equation}
\begin{aligned}
\mathcal{H}=&-\frac{1}{2}f(r)^{-1}  \mathcal{E}^2  +\frac{1}{2}f(r) \left(\frac{\partial S_r}{\partial r}\right)^2+
\frac{1}{2r^2}\left(\frac{\partial S_\theta}{\partial \theta}\right)^2\\
&+\frac{1}{2r^2 \sin ^2 \theta} \left(\mathcal{L}-r^2 \mathcal{B} \sin ^2\theta \right)^2+\frac{1}{2} =0,\label{eq:9}
\end{aligned}
\end{equation}
 where $\mathcal{E} =E/m$, $\mathcal{L}=L/m$ and $\mathcal{B}=\frac{q B}{2 m}$. 
For motion around a magnetized black hole, two conserved quantities associated with the Killing vectors are:
\begin{equation}
\begin{aligned}\label{EL}
\mathcal{E} & = f(r )\dot{t}, \\
\mathcal{L} & = r^2 \sin ^2 \theta  \dot{\phi}+\mathcal{B} r^2 \sin ^2 \theta.
\end{aligned}
\end{equation}
 Eq. \ref{eq:9} can be separated into dynamical and potential parts, namely $\mathcal{H} = H_{\mathrm{dyn}} + H_{pot}$, with
\begin{equation}
\begin{aligned}
H_{\mathrm{dyn}} & =\frac{1}{2}\left[f(r) \left(\frac{\partial S_r}{\partial r}\right)^2+
\frac{1}{r^2}\left(\frac{\partial S_\theta}{\partial \theta}\right)^2\right], \\
H_{pot } & =\frac{1}{2}\left[-\frac{\mathcal{E}^2}{f(r)}+\frac{\left(\mathcal{L}-r^2 \mathcal{B} \sin ^2\theta \right)^2}{r^2 \sin ^2 \theta}+1\right] .\label{eq:10}
\end{aligned}
\end{equation}
For the purpose of further analysis, we shall restrict the motion of the charged test particle to the equatorial plane $( \theta = \pi/2)$. The equation for  the time $t$ the radial motion $r$ and the azimuthal angle $\phi$ of charged particles can be found as \cite{Frolov:2010mi,kolovs2015quasi} 
\begin{equation}
\begin{aligned}
\dot{t} =\frac{\mathcal{E}}{f(r)},  ~~
\dot r ^2 = \mathcal{E}^2 - V_{\text{eff}}(r), ~~
\dot{\phi}=\frac{\mathcal{L}}{r^2}-\mathcal{B}, \label{eq:11}
\end{aligned}
\end{equation}
where the dot in the above equations denotes the derivative with respect to the proper time $\tau$. The effective potential for the radial motion of charged particles is given by: 
\begin{equation}
V_{\text{eff}}(r) =
f(r)\left[1 + \left(\frac{\mathcal{L}}{r}-\mathcal{B} r\right)^2\right].\label{eq:12}
\end{equation}
In this system, we have the main physical quantities $(\mathcal{B}, r, M, g, \mathcal{L})$ which will be rescaled by $M$ into the dimensionless quantities $(\mathcal{B}/M, r/M,  g/M^{2/3}, \mathcal{L}/M)$. Thus, in our theoretical calculations, we shall evaluate only the dimensionless quantities whose physics should not depend on $M$, and we will then set $M=1$ for convenience. As illustrated in FIG. \ref{figa3}, we present the effective potential $V_{\text{eff}}$ as a function of $r$.
From the left panel of FIG. \ref{figa3}, we observe that as the magnetic field parameter $\mathcal{B}$ increases, the extremal values of the effective potential decrease. In the right panel of FIG. \ref{figa3}, we also depict the effective potential $V_{\text{eff}}$ for different values of $g$. It is evident that the effective potential shifts upward as the value of the parameter $g$ increases, predominantly in the range of smaller $r$.

\begin{figure}[ht]
    \centering
    \includegraphics[width=0.4\linewidth]{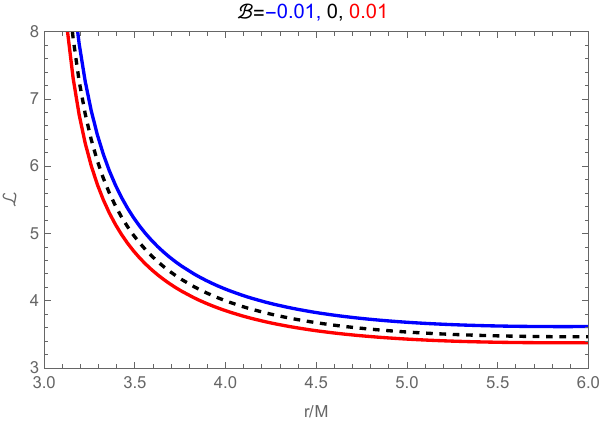}\hspace{1cm}
     \includegraphics[width=0.4\linewidth]{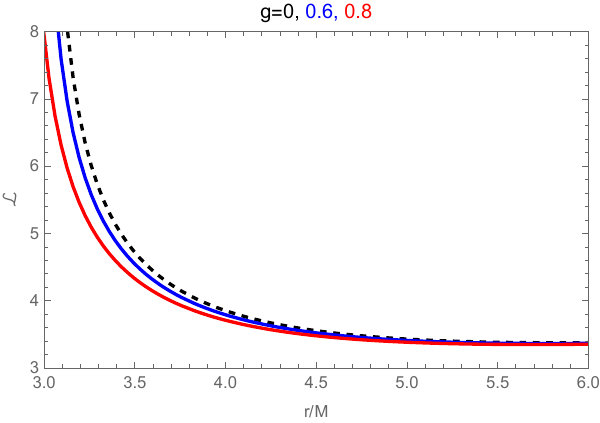}
    \caption{
    Radial dependence of the effective potential for the radial motion of test particles around the magnetized regular black hole. Left panel: $\mathcal{L}$ plots of various combinations of $\mathcal{B}$ with parameter $g=0.1$. Right panel: $\mathcal{L}$ diagram of various combinations of parameter $g$ when fixed magnetic field parameter $\mathcal{B} = 0.01$.}
    \label{figa4}
\end{figure}

To determine the conditions for particles to orbit in circular paths, one needs to solve the following equations simultaneously: 
\begin{equation}
 V_{\text{eff}}(r) = \mathcal{E}^2 ,\quad V_{\text{eff}}'(r) = 0,\label{eq:13}
\end{equation}
where the prime denotes differentiation with respect to $r$. 
From Eq. \ref{eq:13}, we can derive the expression for the angular momentum as follows:
\begin{equation}
\begin{aligned}
\mathcal{L}_{\pm }=\frac{-r^3 \mathcal{B} f'(r)\pm r^{3/2} \sqrt{2 f(r) f'(r)-r f'(r)^2+4 r \mathcal{B}^2 f(r)^2}}{2 f(r)-r f'(r)}.\label{eq:14}
\end{aligned}
\end{equation}
Similarly to Ref.\cite{shaymatov2022constraints}, we focus on the case of $\mathcal{L}_+$, namely $\mathcal{L}\equiv\mathcal{L}_+$.
As shown in FIG. \ref{figa4}, we plot the radial dependence of the $\mathcal{L}$. In the presence of parameters $\mathcal{B}$ and $g$, it is observed that circular orbits are shifted towards smaller $r$, thereby increasing the value of $\mathcal{L}$ for particles on circular orbits with smaller radii.

\section{Epicyclic frequencies for charged particles around magnetized regular black hole}\label{sec:EpicyclicMotion}

When a charged particle is situated in a minimum of the effective potential $V_{\text{eff}}$, corresponding to a stable circular orbit at $r_0$ and $\theta_0 = \pi/2$, if it is slightly perturbed from this stable circular path, the particle will commence oscillating around the minimum, thereby manifesting epicyclic motion governed by linear resonant oscillations.
Given the small perturbations in the radial direction $r = r_0 +\delta r$ and in the latitudinal direction $\theta = \pi/2 +\delta \theta$,
 the equations for linear resonant oscillations can be written as follows \cite{stuchlik2013multi,torok2005radial}:
\begin{equation}
\delta \ddot{r}+\bar{\omega}_r^2 \delta r=0, \quad \delta \ddot{\theta}+\bar{\omega}_\theta^2 \delta \theta=0,\label{eq:15}
\end{equation}
where $\bar{\omega}_r$ and $\bar{\omega}_\theta$ are the radial and latitudinal frequencies for epicyclic oscillations, respectively.

\begin{figure}[ht]
    \centering
      \includegraphics[width=0.45\linewidth]{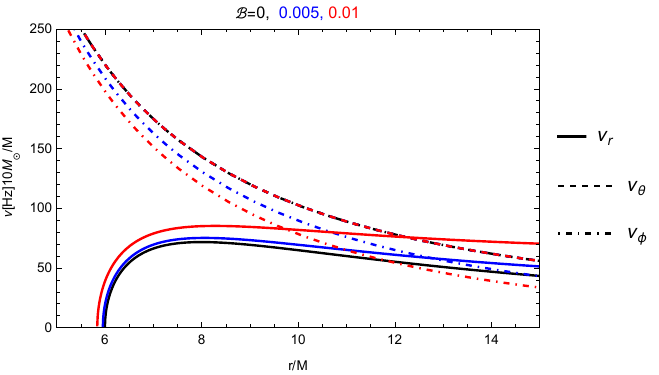}\hspace{1cm}
       \includegraphics[width=0.45\linewidth]{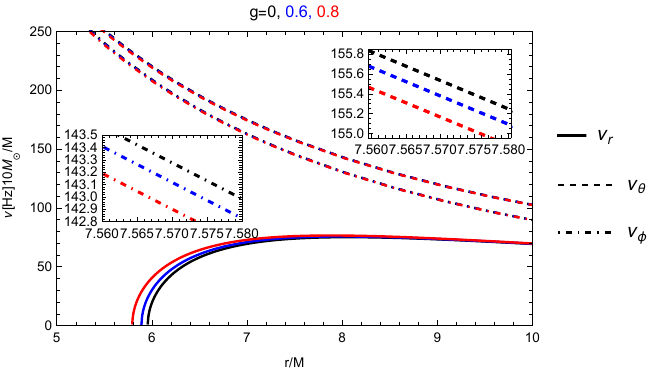}
    \caption{Plot showing the epicyclic frequencies as a function of $r/M$ in the case with parameter $g=0.1$. Radial, latitudinal and orbital frequencies are plotted for various combinations of magnetic field parameter $\mathcal{B}$. Note that solid lines refer to the epicyclic frequencies for the Schwarzschild black hole.}
    \label{figa5}
\end{figure}

These frequencies are measured by a local observer and defined by the following equations \cite{Shaymatov:2020yte,Stuchlik:2021guq}:
\begin{equation}
\begin{aligned}
\bar{\omega}_r^2 & =\frac{1}{g_{r r}} \frac{\partial^2 H_{\mathrm{pot}}}{\partial r^2}, \\
\bar{\omega}_\theta^2 & =\frac{1}{g_{\theta \theta}} \frac{\partial^2 H_{\mathrm{pot}}}{\partial \theta^2}, \\
\bar{\omega}_\phi & =\frac{1}{g_{\phi \phi}}\left(\mathcal{L}-\frac{q}{m} A_\phi\right),\label{eq:16}
\end{aligned}
\end{equation}
where $H_{\mathrm{pot}}$ comes from Eq. \ref{eq:10}. Using the relations Eq. \ref{eq:1}, Eq. \ref{eq:4}, and Eq. \ref{eq:10} in Eq. \ref{eq:16}, we arrive at the formula for the epicyclic frequencies and the azimuthal frequency (as related to the local observers) in the form
\begin{equation}
\begin{aligned}
\bar{\omega}_r^2 &= \frac{\mathcal{E}^2 f''(r)}{2 f(r)}-\frac{\mathcal{E}^2 f'(r)^2}{f(r)^2}+f(r) \left(\frac{3 \mathcal{L}^2}{r^4}+\mathcal{B}^2\right), \\
\bar{\omega}_\theta^2 &= \frac{\mathcal{L}^2}{r^4}-\mathcal{B}^2, \\
\bar{\omega}_\phi &= \frac{\mathcal{L}}{r^2}-\mathcal{B} .\label{pinlv}
\end{aligned}
\end{equation}
The angular frequencies $\bar{\omega}_i$ ($i = r, \theta, \phi$), being locally measured, are pertinent to the observations made by static observers at infinite or very distant distances, who represent the real observers. Therefore, we employ the transformation formula \cite{Shaymatov:2020yte,kolovs2015quasi,Stuchlik:2021guq}:
\begin{equation}
\omega_i= \frac{\bar{\omega_i}}{-g^{t t} \mathcal{E}},\label{eq:17}
\end{equation}
where the locally measured angular frequency is transformed using the redshift factor corresponding to the orbit moving along a stable circular geodesic. The frequencies related to distant static observers, expressed in standard units, are then provided by the relation:
\begin{equation}
\nu_i=\frac{1}{2 \pi} \frac{c^3}{G M} \omega_i.\label{eq:18}
\end{equation}
As depicted in FIG. \ref{figa5}, we present the epicyclic frequencies $\nu_i$ ($i = r, \theta, \phi$) as functions of $r$ for different values of $\mathcal{B}$ and $g$. As illustrated in the left panel of FIG. \ref{figa5}, the radial frequency $\nu_r$ increases and its oscillation shifts to smaller $r$ values with an increase in the magnetic field parameter $\mathcal{B}$. Conversely, the latitudinal frequency $\nu_\theta$ remains almost unchanged, whereas the orbital frequency $\nu_\phi$ experiences a decrease in its value at larger distances, contrasting the behavior around a Schwarzschild black hole. In the right panel of FIG. \ref{figa5}, we depict the radial dependence of the epicyclic frequencies for various values of $g$, with $\mathcal{B}$ held constant. As $g$ increases, the radial frequency shifts to smaller $r$ values, whereas the latitudinal and orbital frequencies exhibit a slight decrease.

\section{Constraints on the magnetic field and parameter }\label{sec:constraint}

In this section, the theoretical results and experimental data of QPOs will be utilized to impose observational constraints on the parameters of magnetized regular black holes.
In particular, we select 4 processed QPOs events from different X-ray binaries: the X-ray observations of GRO J1655-40, XTE J1550-564, XTE J1859+226 and GRS 1919-105 for their QPO frequencies are shown in TABLE \ref{Table1}.
Furthermore, we introduce the periastron precession frequency, $\nu_{per}$, and the nodal precession frequencies, $\nu_{nod}$, as defined in \cite{stella1999correlations}. These are given by:
\begin{equation}
\nu_{per} = \nu_\phi-\nu_r, \quad \nu_{nod} = \nu_\phi -\nu_\theta. \label{eq:19}
\end{equation}
We employ the MCMC simulation method to explore the space of plausible physical parameters and to constrain the range of the parameter $g$.

\subsection{Analysis of Monte Carlo Markov chain }
In this subsection, we analyze the MCMC implementation by $emcee$ \cite{foreman2013emcee} to derive constraints on the magnetized regular black hole with Minkowski core. According to the theorem of Bayes, the posterior probability of a model parameter $(\Theta)$ based on the observed data $(\mathcal{D})$ can be expressed as:
\begin{equation}
\mathcal{P}(\Theta \mid \mathcal{D})=\frac{P(\mathcal{D} \mid \Theta) P(\Theta )}{P(\mathcal{D} )},\label{eq:20}
\end{equation}
 where $P(\mathcal{D} \mid \Theta)$ represents the likelihood of the data given by the model, and $P(\Theta)$ denotes the prior distribution on the parameters, and $ P(\mathcal{D} )$ (called the evidence) is a normalization factor.
In our case, $\mathcal{D}$ represents the QPO frequencies for each X-ray binary, while $\Theta$ represents the astrophysical parameters involved in the QPOs events. The priors of  parameters are assigned as Gaussian distributions within specified limits:
\begin{equation}
      P\left(\theta_i\right) \sim \exp \left[\frac{1}{2}\left(\frac{\theta_i-\theta_{o, i}}{\sigma_i}\right)^2\right], \theta_{\mathrm{low}, i}<\theta_i<\theta_{\mathrm{high}, i},\theta_i=\left[\mathcal{B},r, M\right],
 \end{equation}
with $\sigma_i$ being their respective standard deviations. For parameter $g $, we give the $g \in\left[g_{\text{low}}, g_{\text{high}}\right]$ assign a uniform prior distribution:
\begin{equation}
   P(g) = 
   \begin{cases} 
   \frac{1}{g_{\text{high}}-g_{\text{low}}} & \text{if } g_{\text{low}} \leq g \leq g_{\text{high}} ,\\ 
   0 & \text{otherwise}.
   \end{cases}
\end{equation}

\begin{table}[h]
  {\centering
  \caption{Observational data of orbital frequencies, periastron precession frequencies, and nodal precession frequencies of X-ray binary QPOs.}
  \label{Table1}
  \begin{tabular}{ccccc}
    \toprule
    \hline
     & GRO J1655-40 & XTE J1550-564 & XTE J1859+226 & GRS 1915+105\\
     \midrule
    \hline
         $M(M_\odot)$& 5.4$\pm$0.3\cite{motta2014precise} & 9.1$\pm$0.61\cite{remillard2002evidence,orosz2011improved}& 7.85$\pm$0.46\cite{motta2022black}& 12.4$^{+2.0}_{-1..8}$\cite{remillard2006x}\\
    $\nu_\phi(HZ)$ & 441$\pm$2\cite{motta2014precise}& 276$\pm$3\cite{remillard2002evidence}& 227.5$^{+2.1}_{-2.4}$\cite{motta2022black}& 168$\pm$3\cite{remillard2006x}\\
    $\nu_{per}(HZ)$ & 298$\pm$4\cite{motta2014precise}& 184$\pm$5\cite{remillard2002evidence}& 128.6$^{+1.6}_{-1.8}$\cite{motta2022black}& 113$\pm$5\cite{remillard2006x}\\
    $\nu_{nod}(HZ)$ & 17.3$\pm$0.1\cite{motta2014precise}& - & 3.65$\pm$0.01\cite{motta2022black}& -\\
    \hline
    \bottomrule
  \end{tabular}
  }
\end{table}

We utilize three distinct datasets following the orbital frequencies $\nu_\phi$, periastron precession frequencies $\nu_{\rm per}$, and nodal precession frequencies $\nu_{\rm nod}$. Consequently, the likelihood function $\mathcal{L}$ can be formulated as follows \cite{liu2023constraints,bambi2228testing}:
\begin{equation}
\log \mathcal{L}=\log \mathcal{L}_{obt }+\log \mathcal{L}_{per }+\log \mathcal{L}_{nod },\label{eq:21}
\end{equation}
with
\begin{equation}
\begin{aligned}\label{eq:22}
\log \mathcal{L}_{obt}&=-\frac{1}{2} \sum_i \frac{\left(\nu_{\phi, \mathrm{obs}}^i-\nu_{\phi,  th }^i\right)^2}{\left(\sigma_{\phi, \mathrm{obs}}^i\right)^2}, \\
\log \mathcal{L}_{per }&=-\frac{1}{2} \sum_i \frac{\left(\nu_{per,obs }^i-\nu_{per,th }^i\right)^2}{\left(\sigma_{per,obs }^i\right)^2}, \\
\log \mathcal{L}_{nod }&=-\frac{1}{2} \sum_i \frac{\left(\nu_{nod,obs }^i-\nu_{nod ,  th }^i\right)^2}{\left(\sigma_{nod,obs }^i\right)^2}.
\end{aligned}
\end{equation}
In these equations, the subscript `obs' denotes the corresponding observed quantities while the subscript `th' denotes the corresponding quantities evaluated by the theoretical setup in section \ref{sec:EpicyclicMotion}.
Additionally, $\sigma_{\phi, obs }^i, \sigma_{per, obs}^i$ and $\sigma_{nod,obs }^i$ refer to the statistical uncertainties associated with the related observed quantities.

We employ MCMC techniques to constrain the values of the parameters $\left\{\mathcal{B}, r, M, g \right\}$ for a magnetized regular black hole. Based on the observational data presented in TABLE \ref{Table1}, we initially fit the parameters of a magnetized Schwarzschild black hole ($g=0$) to derive the ranges of $\mathcal{B}, r$, and $M$ at the $68\%$ confidence level(C.L.) , as listed in TABLE \ref{Table2}, serving as priors for the magnetized regular black hole.
Based on these priors in TABLE \ref{Table2} (the range for $g$ is set to $[0,1]$), we randomly generate $10^5$ points for each parameter $\left\{\mathcal{B},r, M,g \right\}$, exploring the physically possible parameter space within the defined boundaries. Finally, using Bayes formula Eq. \ref{eq:20} and the likelihood function Eq. \ref{eq:21}, we apply the MCMC method to obtain the best-fit values for the parameters, as presented in TABLE \ref{Table3}.

\begin{table}[h]
  \centering
  \caption{The Gaussian prior of the magnetized  regular black hole from QPOs for the X-ray Binaries.}
  \label{Table2}
  \begin{tabular}{ccccc}
    \toprule
    \hline
    \multirow{2}{*}{Parameters} & GRO J1655-40 & XTE J1550-564 & XTE J1859+226 & GRS 1915+105  \\
    & $\mu\quad\sigma$ & $\mu\quad\sigma$ & $\mu\quad\sigma$ & $\mu\quad\sigma$  \\
    \midrule
    \hline
    $\mathcal{B}$& -0.00314 0.00003& -0.02009 0.002& -0.00105 0.00001& -0.00725 0.00426 \\
    $r$& 6.71  0.023& 6.30  0.073& 7.42  0.032& 6.59  0.11 \\
    $M(M_\odot)$& 4.41  0.027& 9.15  0.273& 7.19  0.068& 12.16  0.69 \\
    $g$ & Uniform[0,1]& Uniform[0,1]& Uniform[0,1]& Uniform[0,1]\\

    \hline
    \bottomrule
  \end{tabular}
\end{table}

\begin{table}[h]
  \centering
  \caption{The best-fit values of the regular magnetized black hole parameters from QPOs for the X-ray Binaries.}
  \label{Table3}
  \begin{tabular}{ccccc}
    \toprule
    \hline
    Parameters & GRO J1655-40 & XTE J1550-564 & XTE J1859+226 & GRS 1915+105 \\
    \midrule
    \hline
    $\mathcal{B}$ & -0.00314$^{+0.00003}_{-0.00003}$& -0.01985$^{+0.00249}_{-0.00250}$& -0.00105$^{+0.00002}_{-0.00002}$& -0.00681$^{+0.00538}_{-0.00489}$\\
    $r/M$ & 6.70$^{+0.028}_{-0.029}$& 6.29$^{+0.09}_{-0.09}$& 7.41$^{+0.04}_{-0.04}$& 6.61$^{+0.15}_{-0.14}$\\
    $M(M_\odot)$& 4.41$^{+0.030}_{-0.030}$& 9.13$^{+0.28}_{-0.27}$& 7.20$^{+0.07}_{-0.07}$& 12.31$^{+0.79}_{-0.79}$\\
    $g$ & $<0.7014$& $<0.8902$& $<0.7727$& $<0.9441$\\

    \bottomrule
      \hline
  \end{tabular}
\end{table}

\begin{figure}[ht]
    \centering
     \subfigure[~GRO J1655-40]
{\includegraphics[width=0.45\linewidth]{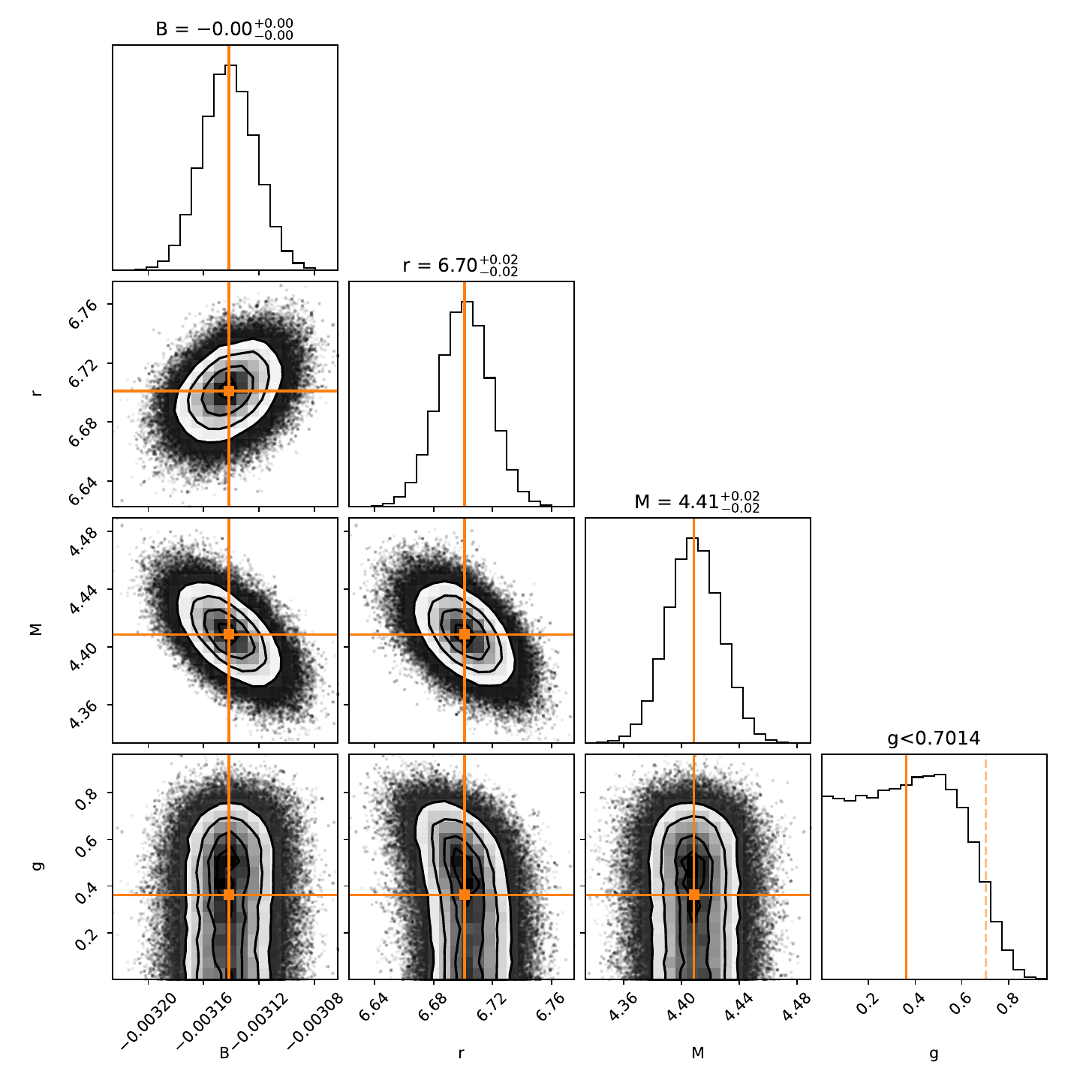}}
\subfigure[~XTE J1550-564]
{\includegraphics[width=0.45\linewidth]{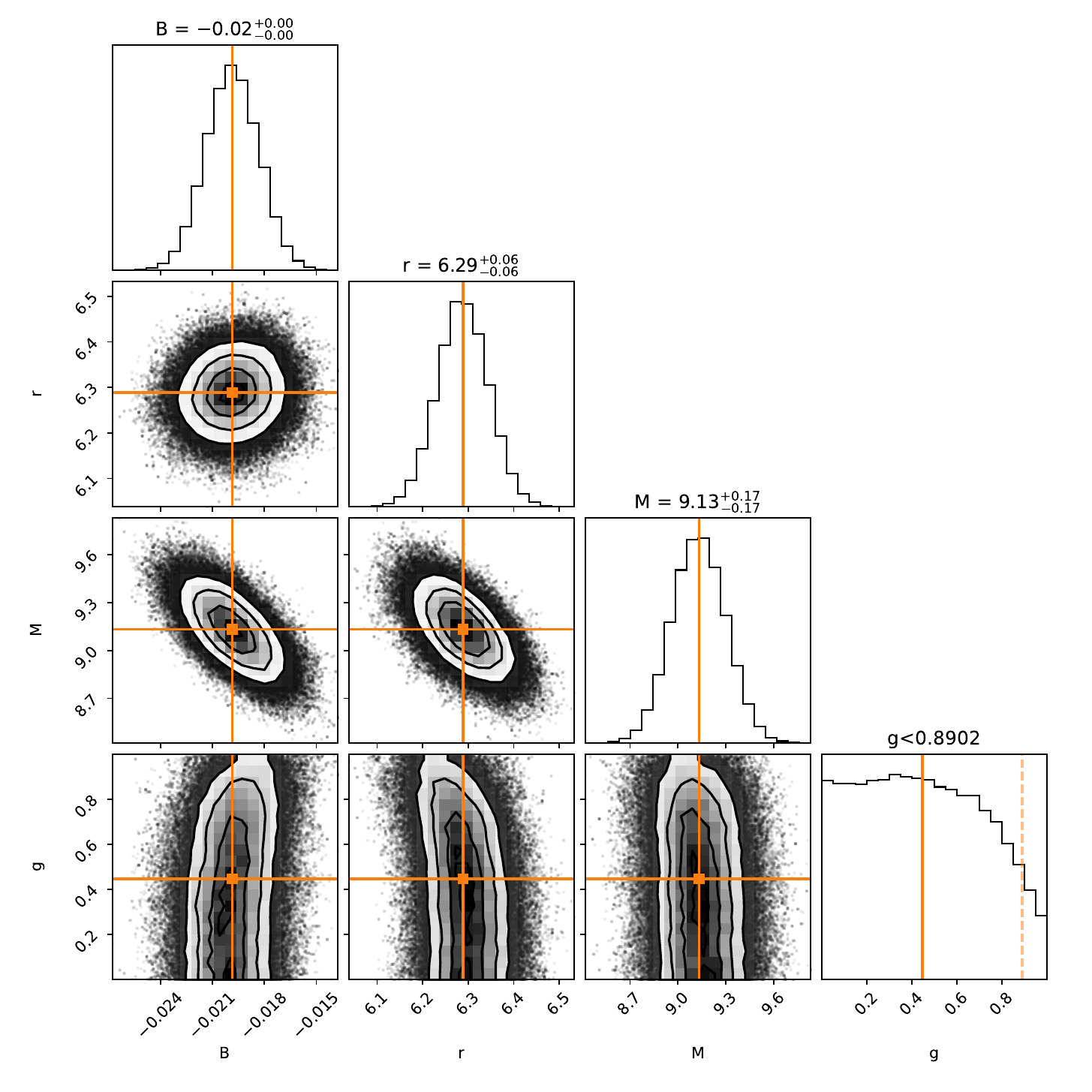}}\\
\subfigure[~XTE J1859+226]
{\includegraphics[width=0.45\linewidth]{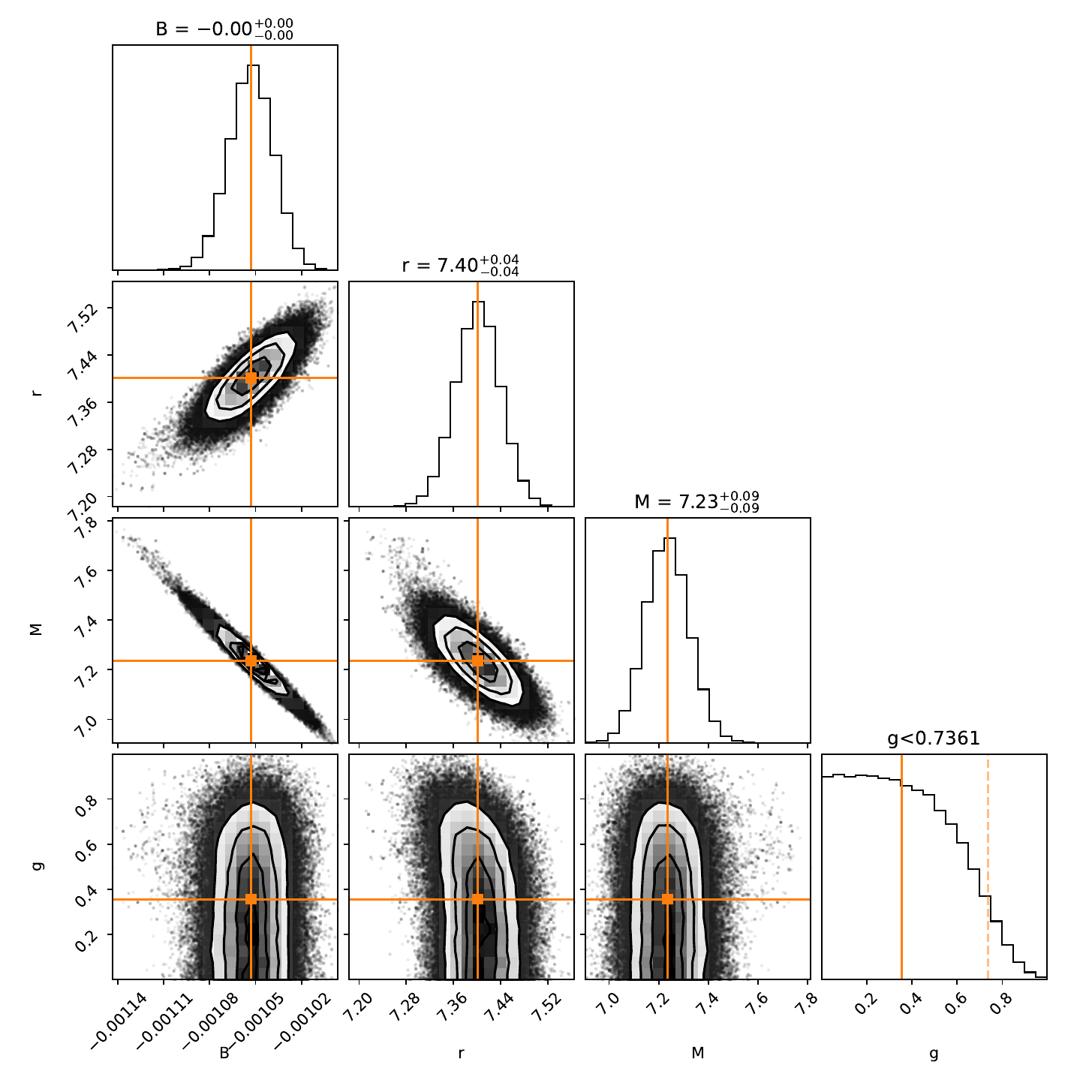}}
\subfigure[GRS 1915+105]
{\includegraphics[width=0.45\linewidth]{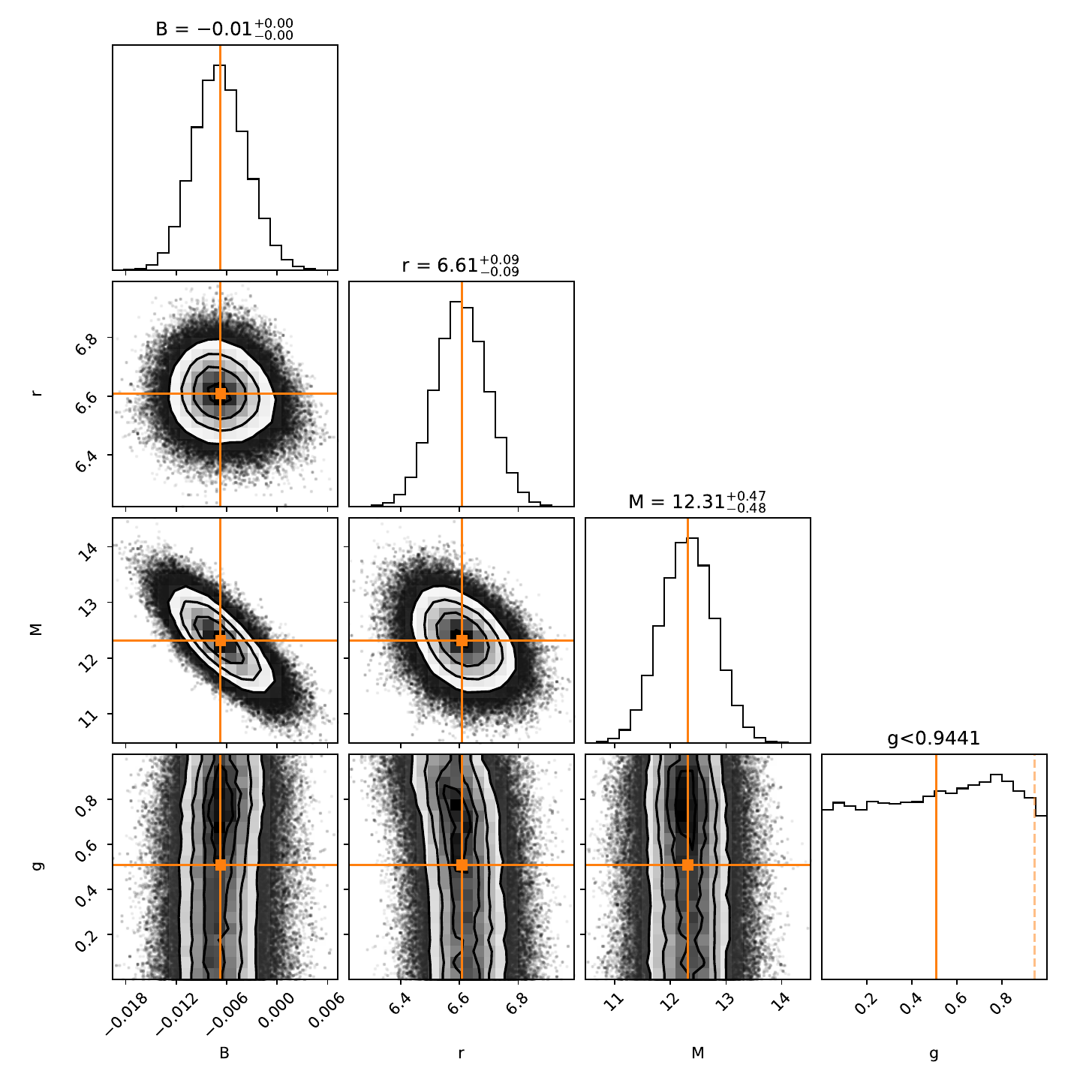}}
    \caption{Constraints on the parameters of the magnetized regular black hole with GRO J1655-40, XTE J1550-
564,  XTE J1859+226 and  GRS 1915+105 from current observations of QPOs within the relativistic precession model.}
    \label{figa9}
\end{figure}

\subsection{Results and Discussions}
Using the setup described in the previous subsections, we explore the 4-dimensional parameter space for the magnetized regular black hole through an MCMC analysis. The best-fit values for these four parameters are presented in TABLE \ref{Table3}. FIG. \ref{figa9} displays the MCMC analysis results for all parameters of the magnetized regular black hole in GRO J1655-40, XTE J1550-564, XTE J1859+226, and GRS 1915+105 respectively. In the contour plots, the shaded regions represent the $68\%$, $90\%$, and $95\%$ C.L. The orange dots represent the best fit points for the entire set of parameters.

The magnetic field strengths $B$ for all four systems, shown in TABLE \ref{Table3}, range from $-0.00105$ to $-0.01985$, indicating relatively low negative values. The weakest magnetic field appears in XTE J1859+226, suggesting that magnetic field effects in this system can be ignored. In contrast, XTE J1550-564 exhibits the strongest magnetic field, implying a potential influence on the dynamics of charged particle orbits. The magnetic field’s impact on QPOs is primarily reflected in slight perturbations to the orbits of charged particles. In systems with weak magnetic fields, QPOs are mainly governed by the gravitational field, with the orbital oscillations of charged particles depending largely on the mass distribution and gravitational properties of the black hole. However, in systems with stronger magnetic fields, such as XTE J1550-564, the magnetic field may cause minor deviations in particle orbits, modulating the QPO frequency.

TABLE \ref{Table3} also shows that the $r$ ratios for the four systems range from $6.29$ to $7.41$, indicating that QPO phenomena occur relatively close to the black hole. For instance, the $r=7.41$ for XTE J1859+226 suggests that QPOs may arise farther from the event horizon, while $r=6.70$ for GRO J1655-40 indicates oscillations closer to the event horizon. This range of characteristic radii implies that these QPOs may be due to orbital oscillations near the black hole’s event horizon, consistent with the steady-state orbital resonance model, where charged particles exhibit quasi-periodic motion along stable orbits in a strong gravitational field. Smaller $r$ values imply that particles are located deeper in the black hole’s gravitational field and experience stronger gravitational effects.

The upper limits of the $g$ parameter for the four systems range from $<0.7014$ (GRO J1655-40) to $0.9441$ (GRS 1915+105). The $g$ value approaching 1 for GRS 1915+105 suggests the presence of significant non-singularity corrections, while the lower $g$ value for GRO J1655-40 indicates that it closely resembles a classical Schwarzschild black hole. The larger $g$ values may signal stronger quantum gravitational effects, particularly in regions of extreme gravitational potential. For GRS 1915+105, the higher $g$ value hints at quantum effects that avoid singularities, whereas systems with smaller $g$ values, such as GRO J1655-40, may align more with classical general relativity predictions, making them ideal candidates for testing between classical and regular black holes. Consequently, the best-fit values of $g$ come from the GRO J1655-40, which gives the upper limit of $g$ is 0.7014 at $95\%$ C.L.

\section{Conclusion}\label{sec:conclusion}

In this paper, we conducted a study on QPOs in X-ray binaries, with a particular focus on the epicyclic motion of charged particles in the vicinity of regular black holes featuring a Minkowski core, immersed in a uniform magnetic field. The effects of magnetic field and regular black hole parameter on the radial effective potential and angular momentum of charged particles are significant. Consequently, they were also found to have prints on the characterized frequencies of the epicyclic oscillations of the particles' circular orbit, which closely connect with the QPOs phenomena in astrophysics.

Then, accompanied by our theoretical results, we fitted the observational QPO data from four X-ray binary systems: GRO J1655-40, XTE J1550-564, XTE J1859+226 and GRS 1915+105. The key parameters constrained through this analysis include the magnetic field strength $B$, the characteristic radius $r$, the mass $M$, and the regular black hole parameter $g$. Our analysis utilized a MCMC method to explore the parameter space, providing the best-fit values for the parameters that describe the QPOs. The QPO signals are highly sensitive to the parameters of black hole, especially to the value of $g$. The MCMC analysis show that the upper limit of the regular black hole correction parameter $g$ is around $0.7014$. Since the parameter $g$ describes the degree of deviation from the classical Schwarzschild black hole, so the constraints from various QPOs observations suggest that regular black holes in current model might deviate from the classical singularity structure of Schwarzschild black holes, and exhibit quantum corrections near the core. 

Our findings could be significant for advancing our understanding of quantum gravity effects near black hole horizons and improving constraints on black hole parameters from astrophysical observations. Further observations with higher precision could refine these constraints, offering deeper insights into the quantum nature of black holes. Moreover, it will be interesting to perform the data simulations with other astronomical signals, such as black hole shadows, and gravitational waves, and further constrain the degree of quantum correction in the current model.

\section*{Acknowledgments}

We are very grateful to Prof. Yi Ling and Prof. Xin Wu for helpful discussions. 
This work is partly supported by the Natural Science Foundation of China under Grants No. 12375054 and 12405067.
It is also supported by the financial support from Brazilian agencies Funda\c{c}\~ao de Amparo \`a Pesquisa do Estado de S\~ao Paulo (FAPESP), Funda\c{c}\~ao de Amparo \`a Pesquisa do Estado do Rio de Janeiro (FAPERJ), Conselho Nacional de Desenvolvimento Cient\'{\i}fico e Tecnol\'ogico (CNPq), and Coordena\c{c}\~ao de Aperfei\c{c}oamento de Pessoal de N\'ivel Superior (CAPES).

\bibliographystyle{utphys}
\bibliography{refers}

\end{document}